\documentclass[a4paper,aps,prl,twocolumn,showpacs,preprintnumbers,amsmath,amssymb]{revtex4}
\usepackage{bbm}
\usepackage{amsmath}
\usepackage{verbatim}
\usepackage{graphicx}
\usepackage{color}
\usepackage{ulem}
\bibliography{References}
\newcommand{\be}{\begin{equation}}
\newcommand{\bea}{\begin{eqnarray}}
\newcommand{\eea}{\end{eqnarray}}
\newcommand{\ee}{\end{equation}}

\def\one{\ensuremath{\hbox{$\mathrm I$\kern-.6em$\mathrm 1$}}}

\def\qed{\leavevmode\unskip\penalty9999 \hbox{}\nobreak\hfill
     \quad\hbox{\leavevmode  \hbox to.77778em{%
               \hfil\vrule   \vbox to.675em%
               {\hrule width.6em\vfil\hrule}\vrule\hfil}}
     \par\vskip3pt}
\newcommand{\beaa}{\begin{eqnarray*}}
\newcommand{\eeaa}{\end{eqnarray*}}
\newcommand{\bma}{\begin{subequations}}
\newcommand{\ema}{\end{subequations}}

\def\one{{\bf 1}}

\def\noxrightarrow[#1]{\dodoublegroupempty\dodoxrightarrow{#1}}
\def\noxleftarrow [#1]{\dodoublegroupempty\dodoxleftarrow {#1}}
\def\dodoxrightarrow#1#2{\mathrel{{\domthxarr0359\rightarrowfill{#1}{#2}}}}
\def\dodoxleftarrow#1#2{\mathrel{{\domthxarr3095\leftarrowfill{#1}{#2}}}}

\begin{document}

\title{\bf Entanglement generated by dissipation and steady state entanglement of two macroscopic objects }

\author{Hanna Krauter$^{1*}$, Christine A. Muschik$^{2*}$, Kasper Jensen$^{1}$, Wojciech Wasilewski$^{1,\dagger}$,  Jonas M. Petersen$^{1}$, J. Ignacio Cirac$^{2}$, and Eugene S. Polzik$^{1,\ddagger}$.}

\affiliation{$^1$ Niels Bohr Institute, Danish Quantum Optics
Center Â QUANTOP, Copenhagen University, Blegdamsvej 17, 2100
Copenhagen Denmark.\\ $^2$Max-Planck--Institut f\"ur Quantenoptik,
Hans-Kopfermann-Strasse, D-85748 Garching, Germany\\
$^*$These authors contributed equally to this work.}
\begin{abstract}
Entanglement is a striking feature of quantum mechanics and an
essential ingredient in most applications in quantum information.
Typically, coupling of a system to an environment inhibits
entanglement, particularly in macroscopic systems. Here we report
on an experiment, where dissipation continuously generates
entanglement between two macroscopic objects. This is achieved by
engineering the dissipation using laser- and magnetic fields, and
leads to robust event-ready entanglement maintained for $0.04$s at
room temperature. Our system consists of two ensembles containing
about $10^{12}$ atoms and separated by $0.5$m coupled to the
environment composed of the vacuum modes of the electromagnetic
field. By combining the dissipative mechanism with a continuous
measurement, steady state entanglement is continuously generated
and observed for up to an hour.
\end{abstract}


\maketitle
To date, experiments investigating quantum superpositions and
entanglement are hampered by decoherence. Its effects have been
studied in several systems \cite{RefSet1}. However, it was
recognized \cite{PoyatosCiracZoller} that the engineered
interaction with a reservoir can drive the system into a desired
steady state. In particular, dissipation common for two systems
can drive them into an entangled state \cite{RefSet2}.
The idea of using and engineering dissipation rather than relying
on coherent evolutions only, represents a paradigm shift with
potentially significant practical advantages. Contrary to other
methods, entanglement generation by dissipation does not require
the preparation of a system in a particular input state and
exists, in principle, for an arbitrary long time, which is
expected to play an important role in quantum information
protocols \cite{DLCZ,Kimble08,RMP2010,DissipativeRepeater}. These
features make dissipative methods inherently stable against weak
random perturbations, with the dissipative dynamics stabilizing
the entanglement.

We report on the first demonstration of purely dissipative
entanglement generation \cite{Preprint}. In contrast to previous
approaches
\cite{Polzik2001,KuzmichKimbleeisaman2005yuan2008,appel2009},
entanglement is obtained without using measurements on the quantum
state of the environment (i.e. the light field). The
dissipation-based method implemented here is deterministic and
unconditional and therefore fundamentally different from standard
approaches such as the QND-based method \cite{Polzik2001} or the
DLCZ protocol \cite{DLCZ}, which yield a separable state if the
emitted photons are not detected. Furthermore, we report the
creation of a steady state atomic entanglement by combining the
dissipative mechanism proposed in \cite{TheoryPaper} with
continuous measurements. The generated entanglement is of the EPR
type, which plays a central role in continuous variable quantum
information processing \cite{Furusawa2007,RMP2010}, quantum
sensing \cite{magnetometry} and metrology
\cite{appel2009,Vuletic,OberthalerRiedel2010}.

Fig.~1a presents the principles of engineered dissipation in our
system consisting of two $^{133}$Cs ensembles, interacting with a
$y$-polarized laser field at $\omega_L$. A pair of two-level
systems is encoded in the $6S_{1/2}$ ground state sublevels
$|\!\!\uparrow\rangle_{I}\equiv|4, 4\rangle_I$,
$|\!\!\downarrow\rangle_{I}\equiv|4, 3\rangle_I$ and
$|\!\!\uparrow\rangle_{II}\equiv|4, -3\rangle_{II}$,
$|\!\!\downarrow\rangle_{II}\equiv|4, -4\rangle_{II}$. Operators
$J^\pm_{\text{I/II}}$ with
$J^-=\sum_{i=1}^{N}|\!\!\uparrow\rangle_{i}\langle
\downarrow\!\!|$ describe collective spin flips, where $N$ is the
number of atoms. The atoms are placed in a magnetic field in the
$x$-direction and the collective operators
$J_{y}=\sqrt{2}\left(J^+ + J^{-}\right)$ and
$J_{z}=i\sqrt{2}\left(J^+ - J^{-}\right)$ are defined in the frame
rotating at the Larmor frequency $\Omega$. The two ensembles are
initialized by optical pumping along the $x$-axis in the extreme
states $m_F=4$ and  $m_F=-4$ respectively, corresponding to
$\langle J_{x}\rangle\equiv\langle J_{x,I}\rangle=-\langle
J_{x,II}\rangle\approx 4 N$ (see Fig.~1).
Within the Holstein-Primakoff approximation, we introduce the
canonical variables $X_{I/II}= J_{y,I/II}/\sqrt{|\langle
J_{x}\rangle|}$ and $P_{I/II}=\pm J_{z,I/II}/\sqrt{|\langle
J_{x}\rangle|}$ \cite{RMP2010}. The EPR entanglement condition
\cite{DGCZ,Polzik2001} for such ensembles is given by
$\xi=\Sigma_{J}/\left(2|\langle J_{x}\rangle|
\right)=\text{var}(X_I-X_{II})/2+\text{var}(P_I+P_{II})/2<1$,
where $\Sigma_{J}=\text{var} (J_{y,\text{I}}- J_{y,\text{II}})+
\text{var}(J_{\text{z},I} - J_{z,\text{II}})$.
%
\begin{figure}
\includegraphics[width=.45\textwidth]{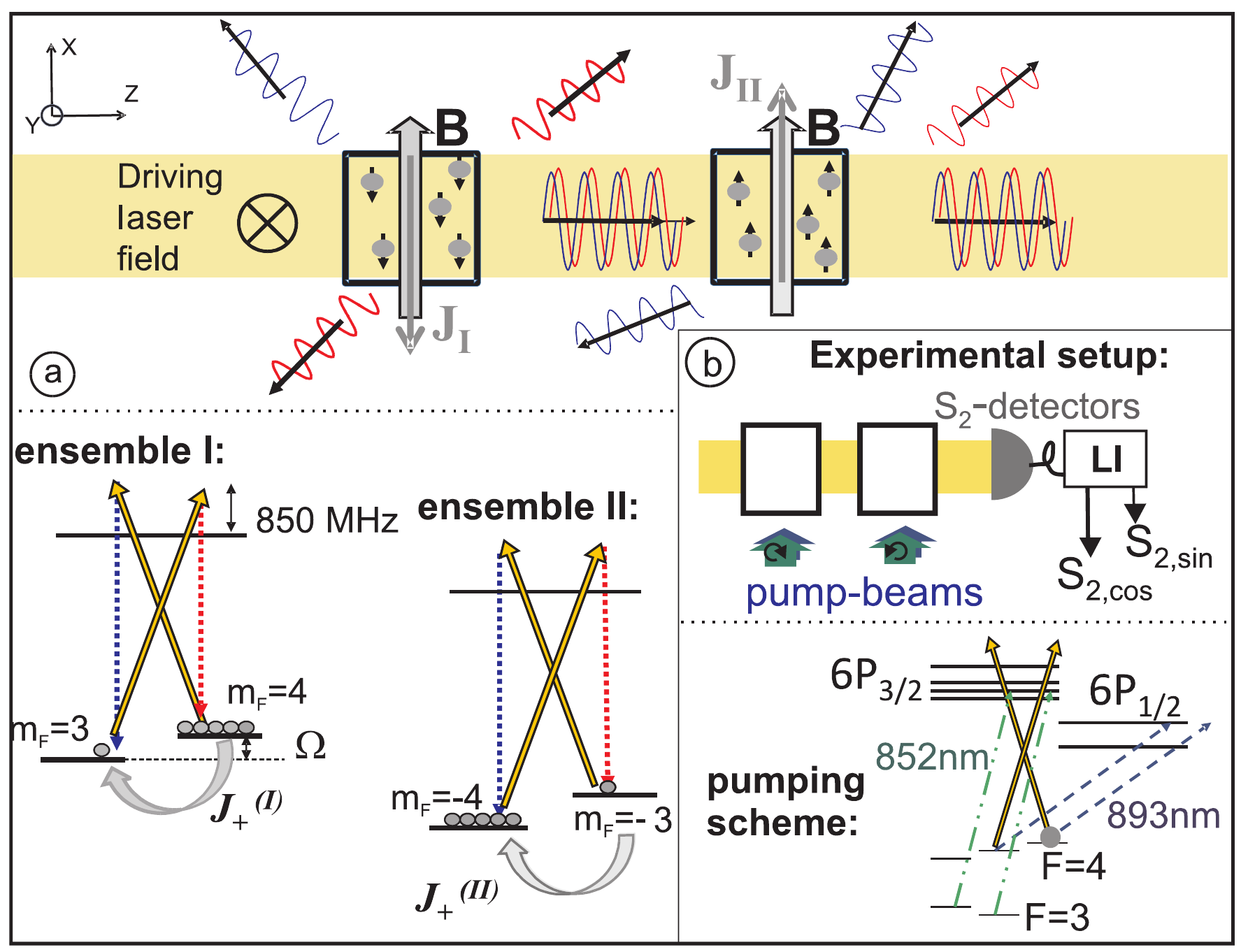}
\caption{a) Collective dissipation modes and atomic levels: two
spatially separated atomic ensembles interact with the environment
composed of the vacuum modes of the electromagnetic field. The
coupling is driven by the ${y}$-polarized laser beam. The
engineered collective dissipation is due to photons scattered in
the forward ${z}$-direction. Internal level scheme of the atoms:
the effective two-level systems $|\!\!\uparrow \rangle$ and
$|\!\!\downarrow\rangle$ are two magnetic sublevels Zeeman-shifted
by a magnetic field applied in the ${x}$-direction, which defines
the quantization axis. Atoms in the two ensembles are initialized
in opposite spin states. The laser beam off--resonantly couples
these levels to the excited states and to the electromagnetic
vacuum modes. Due to the Zeeman shift of the ground state levels,
photons are emitted into the upper and lower sidebands (shown in
blue and red color) leading to collective spin flips $J_{\pm}$. b)
Geometry of the experiment. The  $S_2$ detector signal processed
by the lock-in amplifier (LI) is used to determine the atomic
quantum spin components $J_{y,z}$ as described in the text. The
optical pumping scheme is also shown.}\label{figure1}
\end{figure}
%
The entangling mechanism is due to the coupling to the
$x$-polarized vacuum modes in the propagation direction $z$ of the
laser field (Fig.~1),
which are shared by both ensembles and provide the desired common
environment. Spin flip processes in the two samples accompanied by
forward scattering result in indistinguishable photons leading to
quantum interference and entanglement of the ensembles. These spin
flips and the corresponding photon scattering (see level schemes
in Fig.~1) are described by the interaction Hamiltonian of
the type\\
$ H\propto \int_{\Delta\omega_{\text{ls}}}d\mathbf{k}\left(A
a_\mathbf{k}^{\dag}+A^{\dag}a_\mathbf{k}\right)+\int_{\Delta
\omega_{\text{us}}}d\mathbf{k}\left(B
a_\mathbf{k}^{\dag}+B^{\dag}a_\mathbf{k}\right)
$\\
 where the
integrals cover narrow bandwidths
centered around the lower and upper sideband at
$\omega_L\mp\Omega$ respectively and with the non-local spin
operators $A=\mu J^-_{\text{I}}-\nu J^-_{\text{II}}$, $B=\mu
J^+_{\text{II}}- \nu J^+_{\text{I}}$. The fact that the
electromagnetic modes $a_{\mathbf{k}}^{\dag}$ form a continuum is
crucial for the entanglement to be created without measurements
\cite{TheoryPaper}.
As emission into the forward direction is collectively enhanced
for a large optical depth $d$ \cite{RMP2010}, the forward
scattered modes can successfully compete with spontaneous emission
modes in directions other than $z$ which leads to decoherence of
the atomic state.
 Note that the generation of entanglement cannot
be explained by the interaction of photons emitted by the first
ensemble with the second one, which is negligible in our parameter
regime.
The nonlocal dissipative atomic dynamics obtained after tracing
over the photonic modes is governed by the master equation
\cite{TheoryPaper}:\\
  $\frac{d}{dt}\rho=d\ \frac{\Gamma}{2}\left(A\rho A^{\dag} -A^{\dag} A \rho+B\rho B^{\dag}
 -B^{\dag} B \rho+H.C.\right)
 +{\cal L}_{\rm noise}\rho$,
%
where $\rho$ is the atomic density operator, and $\Gamma$ is the
single atom radiative decay. The Lindblad terms in parentheses,
which would usually describe regular spontaneous emission, drive
the system into an EPR state with $\xi=(\mu-\nu)^{2}<1$
\cite{TheoryPaper}, due to the special nonlocal construction of
$A$ and $B$.  ${\cal L}_{\rm noise}$ describes undesired processes
such as single atom spontaneous emission, collisions, etc.
\begin{figure*}
\includegraphics[width=.9\textwidth]{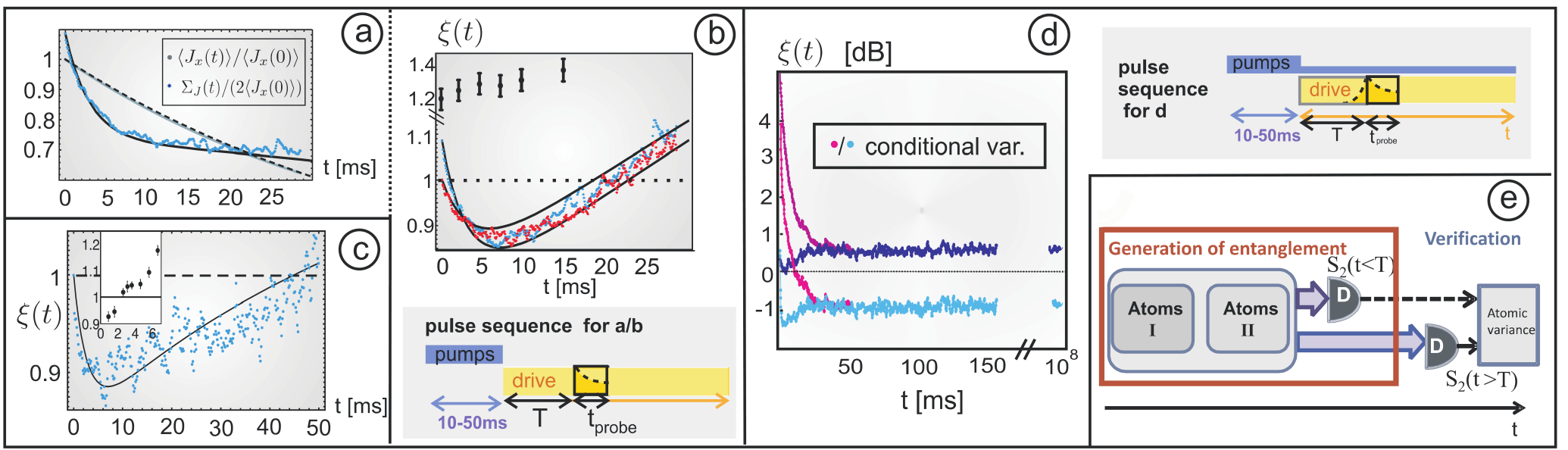}
\caption{Entanglement generated by dissipation (a-c) and steady
state entanglement (d). a) Time evolution of
$\Sigma_{J}(t)/\left(2|\langle J_{x}(0)\rangle|\right)$ (blue) and
$\langle J_{x}(t)\rangle/\langle J_{x}(0)\rangle$ (grey). The
theoretical fits (full and dashed black line) are based on the
parameters $d=55$ (optical density),
$\Gamma_{\text{col}}=0.002$ms$^{-1}$ (collisional rate) and
$\tilde{\Gamma}=0.193$ms$^{-1}$ (dephasing rate \cite{Footnote}).
The rates for driving field induced transitions $|4,\pm
3\rangle\rightarrow |4,\pm 4\rangle$ and $|4,\pm
4\rangle\rightarrow |4,\pm 3\rangle$ are given by $\mu^2\Gamma$
and $\nu^2\Gamma$ respectively, where $\Gamma=0.002$ms$^{-1}$.  b)
Entanglement $\xi(t)$ versus time in ms. Blue data points
correspond to the results shown in a). Data points in orange are
obtained for a lower optical depth ($d=35$). The other parameters
used in the fits take the same values as in a). $\xi(t)<1$
certifies the creation of an inseparable state. The relevant pulse
sequence is shown below. The data taken in the absence of the
driving field (black points) show no entanglement. c) Dissipative
entanglement generation in the presence of the pump field which
incoherently transfers atomic population from undesired levels
within $F=4$ back to the two level subsystem. The pump rate is
$\Gamma_{\text{pump}}=0.168$ms$^{-1}$. $d=37$, the fitting
parameters $\Gamma_{\text{col}}$ and $\Gamma$ take the same values
as in a) and $\tilde{\Gamma}=0.233$ms$^{-1}$. The inset shows the
evolution of $\xi(t)$ after the driving field is switched off
after entanglement is generated by dissipation. d) Entanglement
$\xi(t)$ for different initial conditions. The upper curves show a
purely dissipative evolution. The lower curves - the entanglement
generated by dissipation combined with the measurement. Points on
the right represent an average over measurements of one hour where
atoms were kept in a steady state. The used exponential time mode
functions are depicted in the pulse sequence and are described in
the text. e) Schematic illustration of entanglement generation and
verification. The signal from the detector D for times $t>T$ is
used for verification of entanglement in (a-c). In d) the signal
taken at $t<T$ is given to the verifier as additional
information.} \label{figure3}
\end{figure*}
%
%
%
The experiments are performed using two dilute $^{133}$Cs gas
samples in $2.2$cm cubic cells separated by $0.5$m described
elsewhere \cite{RMP2010}.  A bias magnetic field of $0.9$G leads
to a Zeeman splitting of $\Omega=2\pi\cdot322$kHz (see Fig.~1).
The anti-relaxation coating of the cell walls and careful magnetic
shielding \cite{magnetometry} provide the non-radiative
decoherence time for populations and coherences of
$T_{1}\approx130$ms and $T_{2}\approx40$ms.
The two ensembles are initialized in the states $|4,\pm4\rangle$
with orientation up to $P=0.998(3)$ by applying a pump laser
polarizing the $F=4$ manifold and a laser repumping atoms from
$F=3$ to $F=4$ for $10$ to $50$ms (Fig.~1b). The driving laser is
blue detuned by $850$MHz from the $F=4\leftrightarrow F=5$
transition of the $D_{2}$ line corresponding to
$\left(\mu-\nu\right)^2=0.16$. The laser power influences both the
collective and the single atom dissipation processes and has been
optimized within a range of $5-15$mW.
The nonlocal atomic state variance $\xi=\Sigma_{J}/\left(2|\langle
J_{x}\rangle| \right)$ is inferred, and the entanglement condition
$\xi<1$ is verified by a local polarization measurement on the
light transmitted through the two ensembles (Fig.~1b). We use the
same laser to create and to verify the entanglement which
significantly simplifies the experiment. In the period $t<T$, up
to a variable time $T$ (see the pulse sequence in Fig.~2b) the
laser serves only as the driving source for dissipation. The
results of the measurements on the transmitted light are not used,
which is equivalent to tracing out the light field. Beginning at
$t=T$, the temporal mode of the transmitted light is used for the
determination of the atomic state at time $T$ using the
established method \cite{RMP2010,appel2009,Vuletic,Koschorreck10}
of linear mapping of the atomic state onto light (atomic
tomography via quantum polarization spectroscopy). The particular
linear mapping used here has been utilized in several other
contexts \cite{squeezed,magnetometry,QuantumMemory} and is
described by the input-output relations for atomic and light
operators before and after the interaction:
   \begin{eqnarray}
  \tfrac{1}{\sqrt{2}} ( {X}_I^{\text{out}}\!\!-\!\!{X}_{II}^{\text{out}})\!\!&=&\!\!e^{-\gamma_s T}\cdot\tfrac{1}{\sqrt{2}}( {X}_I^{\text{in}}\!\!-\!\! {X}_{II}^{\text{in}}) - \kappa (\mu\!-\!\nu)^2 \nonumber {y}_{c+}^{\text{in}},\\
   {y}_{c-}^{\text{out}}\!\!&=&\!\!{y}_{c+}^{\text{in}} e^{-\gamma_s T}\!+\!\kappa\cdot\tfrac{1}{\sqrt{2}}( {X}_I^{\text{in}}\!-\! {X}_{II}^{\text{in}}),\label{IOeq}
   \end{eqnarray}
   and similarly for $ {X}_I- {X}_{II} \rightarrow {P}_I+ {P}_{II}$ and ${y}_{c} \rightarrow {y}_{s}$.
Here $\kappa^2=(1-e^{-2\gamma_s T})/(\mu-\nu)^2$ and
$\gamma_s\propto (\mu-\nu)^2 J_x \Phi$, where $\Phi$ is the flux
of photons in the drive field  and $T$ is the interaction time.
The light operators are given by the $\cos(\Omega t)$ and
$\sin(\Omega t)$ components of the Stokes operator $S_2$ weighted
with an exponentially falling (rising) mode function:
${y}_{c-/+}^{\text{out(in)}}=
\tfrac{1}{N_{\pm}\sqrt{S_x}}\int_0^{T}
S_2^{\text{out(in)}}(t)\cos(\Omega t)e^{\mp \gamma_s t}dt$
(analogously for sine modes). Just as the master equation does,
the input-output relations predict an entangled atomic state with
variance $\Sigma_{J}(t) \rightarrow (\mu-\nu)^2 $, for
$T>>\gamma_{s}^{-1}$. For $(\mu-\nu)^2<<1$ and finite
$\gamma_{s}T$ the input-output relations reduce to the
quantum-nondemolition type.
The atomic EPR variance $\Sigma_{J}(T)$ at time $T$ can be
inferred by using the input-output relations (\ref{IOeq}). The
Stokes operator $S_2$ (the photon flux difference between
$+45^{o}$ and $-45^{o}$ polarizations with respect to the $y$-axis
in Fig.~1b) is measured in the time interval
$[T;T+t_\text{probe}]$ (see pulse sequence in Fig.~2b) with the
photocurrent electronically processed to obtain the relevant light
mode: ${y}_{c-}^{\text{read out}}\propto \int_T^{T+t_\text{probe}}
S_2^{\text{out}}(t)\cos(\Omega t )e^{- \gamma_st}dt$. The
parameters $\kappa^2,\gamma_s,(\mu - \nu)$ of the linear
input-output relations are calibrated as described elsewhere
\cite{squeezed,magnetometry}. The atomic state reconstruction is
calibrated and verified carefully as described in detail in the
supplemental material SM \cite{SOM}, where also the modification
of the input output equations by losses and decoherence is
presented. We conclude that
the measurement of $\xi$ is reliable within the uncertainty of
$\pm 4 \%$ arising from uncertainty in the measurements of
$\kappa^2$, the detection efficiency $\eta$ and the shot noise of
light.

In the first set of experiments, entanglement is generated purely
dissipatively. In the first series of this set, the pump- and
repump fields are turned off at time $t=0$ (Fig.~2a,b) and the
driving (entangling) laser is turned on. In the presence of the
drive field  ($P\approx5.6$mW) $T_{2}$ is reduced to $6$ms and
$T_{1}$ to $34$ms. This decoherence has been considered the
fundamental limitation for the entanglement generated by
measurements \cite{RMP2010}. Here, the collective entangling
dissipation due to forward scattering dominates over the single
atom decoherence and leads to a rapid reduction of $\Sigma_{J}(t)$
on the time scale of $\gamma_{s}^{-1}$. Fig.~2a shows the time
evolution of $\Sigma_{J}(t)$ normalized to $2|\langle
J_{x}(0)\rangle|$. For a Coherent Spin State (CSS) $\xi=1$, and
$\Sigma_{\text{CSS}}=2|\langle J_{x}\rangle|$ defines the
projection noise (PN) level, below which lies the noise level of
entangled states. The dynamics of $2|\langle J_{x}(t)\rangle|$ due
to single atom spontaneous emission and collisions on the slow
time scale of $T_{1}$ is also shown in Fig.~2a. Fig.~2b presents
the time evolution of entanglement for two values of the optical
depth $d\approx 34$ ($\Theta=8.5^o$) and $d\approx 56$
($\Theta=14.0^o$). The data is well fitted with theory
\cite{TheoryPaper} using the collisional rate $\Gamma_\text{col}$
and dephasing rate $\tilde\Gamma$ \cite{Footnote} compatible with
experimental values. The details of calculations of the fits are
given in \cite{SOM}. The time interval $0.015$s over which
entanglement is continuously maintained  is several times longer
than the best previous results obtained for measurement induced
entanglement \cite{Kimble08,RMP2010} and much longer than $T_2$.
For comparison, if the driving (entangling) laser is off during
$0<t<T$ and is turned on only at $t=T$ to measure the atomic
state, $\xi(T)$ predictably stays above the PN level (black points
in Fig.~2b). Also a slight mismatch of the Larmor frequencies of
the two ensembles in the preparation period by $\sim20$Hz leads to
the disappearance of the entanglement. This can be viewed as a
direct consequence of the "which way" information due to the
distinguishability of photons emitted by the two ensembles.

In the series presented above, entanglement is created in a quasi
steady state rather than in a steady state, as would be the case
for atoms with a true two-level atomic ground state, for example
in Ytterbium ensembles \cite{Ytterbium}. On the time scale of
$T_{1}$, atoms are lost to other magnetic sublevels of $F=4$ and
to the level $F=3$. This causes the eventual extinction of
entanglement as described well by the theoretical fits shown in
Fig.~2a,b with the pumping rate $\Gamma_{\text{pump}}$ being close
to the experimental value. In the next series of experiments, the
pumping field of an optimal strength resonant with the $F=4$ state
is kept on during the entanglement generation period $t>0$ (
Fig.~2c). Remarkably, this incoherent process does not suppress
generation of entanglement, but on the contrary brings it further
towards a steady state. The entanglement can now be maintained for
$0.04$s thanks to pumping atoms from sublevels $|m_{F}|\leq 3$
which contribute higher noise, back to $|m_{F}|=4$  which is a
dark state for the pump beam. The eventual loss of entanglement,
is in part due to atoms which are lost to the $F=3$ ground state,
effectively reducing $d$. If the entangling mechanism is turned
off, the entangled state decays in $2$ms (inset in Fig.~2c), as
expected \cite{RMP2010} from the decoherence in the dark.

Finally, we demonstrate generation of steady state atomic
entanglement. To this end, a repumping field  $F=3 \rightarrow
F=4$  is added during the entanglement generation, thus closing
down the last escape channel from the relevant spin system. The
atoms reach a steady state which is, however, not entangled
because the collective processes are not sufficiently strong to
overcome the noise added by the incoherent repumping field. Theory
predicts \cite{TheoryPaper} that a steady state entanglement can
be achieved for $d=100$, but this is experimentally unfeasible.
However, we can use the fact that due to single atom decoherence
sources, the atomic state is not pure, and hence forward scattered
light is not completely disentangled from the atoms. Up to now,
measurements on light variables  have only been used to verify
entanglement at time T for which only $y_{c,s-}^{out}(t>T)$ have
been utilized. Using the results of the continuous measurement on
the open atomic quantum system during the interval $t<T$, we can
enhance the entanglement generated by dissipation and maintain it
in the steady state. In this scenario, the verifier (Fig.~2e)
receives the classical information $S_2^{out}(t<T)$ which is used
to calculate the conditional variance
$\text{var}({y}_{c,s}^\text{cond})=\text{var}({y}_{c,s}^{\text{read
out}}(T)-\alpha{y}_{c,s}^\text{feed}(t<T))$. Here,
${y}_{c}^\text{feed}(t<T)= \tfrac{1}{N_f\sqrt{S_x}}\int_0^T
S_2^{\text{out}}(t)\cos(\Omega t)e^{\gamma_m t}dt$ and the
feedback gain $\alpha$ and the time constant $\gamma_m$ are
optimized to achieve maximal noise reduction. The light mode that
brings about the best noise reduction is a fast growing
exponential mode, with
$\gamma_m=0.83\text{ms}^{-1}>\gamma=1/T_2=0.27\text{ms}^{-1}$. The
conditionally reduced atomic variance
$\xi_\text{cond}=\text{var}[\tfrac{1}{2}(X_I-X_{II})+\tfrac{1}{2}(P_I-P_{II})
-\frac{\alpha}{\kappa}({y}_{c}^\text{feed}+{y}_{s}^\text{feed})(t<T))]$
is then found from $\text{var}({y}_{c,s}^\text{cond})$ using the
same calibrated input-output relations as above. These central
results are displayed in Fig.~2d, which shows the evolution of the
variances of the purely dissipatively generated atomic state
(upper curves) and the entanglement produced using the hybrid
method including dissipation and continuous measurements (lower
curves). Each pair of curves is taken with two different initial
conditions. These results demonstrate a very important aspect of
our work, - they show that the generated steady state is
independent of the initial state, and that entanglement is
maintained for up to an hour, if dissipative processes are
combined with measurements.

In conclusion, we have observed entanglement of macroscopic atomic
ensembles generated by dissipation and the steady state atomic
entanglement. The results present a new step in quantum control of
entanglement. Dissipatively generated entanglement provides not
only event-ready entangled links for standard protocols but is
also an elementary resource for future applications in continuous
quantum information processing schemes, such as dissipative
distillation and repeater protocols, which allow for the
distribution of long-range high-quality steady state entanglement
\cite{DissipativeRepeater}.

\section*{Acknowledgements}
We acknowledge support from the Elite Network of Bavaria (ENB)
project QCCC and the EU projects COMPAS, Q-ESSENCE and QUEVADIS.
C.A.M. acknowledges valuable discussions with K. Hammerer, K.G.H.
Vollbrecht and G. Giedke.

\section*{Author Contributions}
H.K. and K.J. obtained the experimental data with the assistance
of J.M.P. W.W. performed preliminary measurements. C.M. developed
the theory. J.I.C. and E.S.P. planned and supervised the work.\\
\\ \small{$^{\dagger}$ Current address: Institute of Experimental
Physics, University of Warsaw, Hoza 69, PL00681 Warsaw, Poland}\\
\small{$^{\ddagger}$ Author for correspondence. E-mail:
polzik@nbi.dk}

%

%
%
%
\newpage
\renewcommand{\figurename}{Supplemental figure}
\setcounter{figure}{0} \setcounter{equation}{0}
\renewcommand{\thefigure}{S.\arabic{figure}}
\renewcommand{\theequation}{S.\arabic{equation}}

\newcommand{\var}{\text{var}}

\newpage

\section*{SUPPLEMENTAL MATERIAL}

The first part of the supplemental material contains details of
the calibration of the atomic spin noise with respect to the
projection noise and of the orientation of the collective atomic
spin.\par The second part contains details of the theoretical fits
presented in Fig.2 of the main text.
\section*{Projection noise calibration}
The variances of the collective atomic operators
$P_I^\text{in}+P_{II}^\text{in}$ and
$X_I^\text{in}-X_{II}^\text{in}$ are found from the measured
polarization mode of the transmitted light $y_{c,s-}$ using the
input-output relations given in Eq.~(1) in the main text, as
follows:
\begin{eqnarray}
\var(P_I^\text{in}+P_{II}^\text{in})/2=(\var(y_{c-}^\text{out})-\sigma_{in}^2(1-\kappa^2(\mu-\nu)^2))/\kappa^2,\nonumber\\
\var(X_I^\text{in}-X_{II}^\text{in})/2=(\var(y_{s-}^\text{out})-\sigma_{in}^2(1-\kappa^2(\mu-\nu)^2))/\kappa^2,\label{noiserec}
\end{eqnarray}
where  $\sigma_{in}^2$ is the shot noise of light. The normalized
EPR variance of atomic noise is
$\xi=\var(P_I^\text{in}+P_{II}^\text{in})/2+\var(X_I^\text{in}-X_{II}^\text{in})/2$.
\par

\begin{figure}
\includegraphics[width=.4\textwidth]{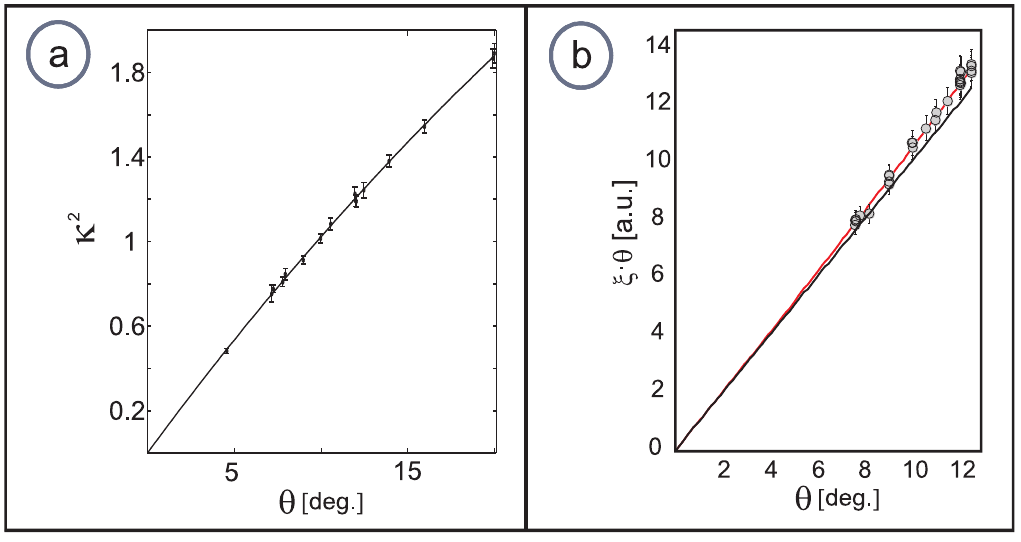}
\caption{\textbf{Calibration measurements:} a) Normalized initial
spin noise variance as a function of $\Theta \propto J_x$. The
light power was held constant at $5.6$mW, the pulse duration was
$1$ms. Red line - linear plus quadratic fit. Black line - linear
part of the fit presenting the projection noise. b) Measurement
results for $\kappa^2$ for different Faraday angles $\theta$.}
\label{fig:calibB}
\end{figure}

\begin{figure}
\includegraphics[width=.35\textwidth]{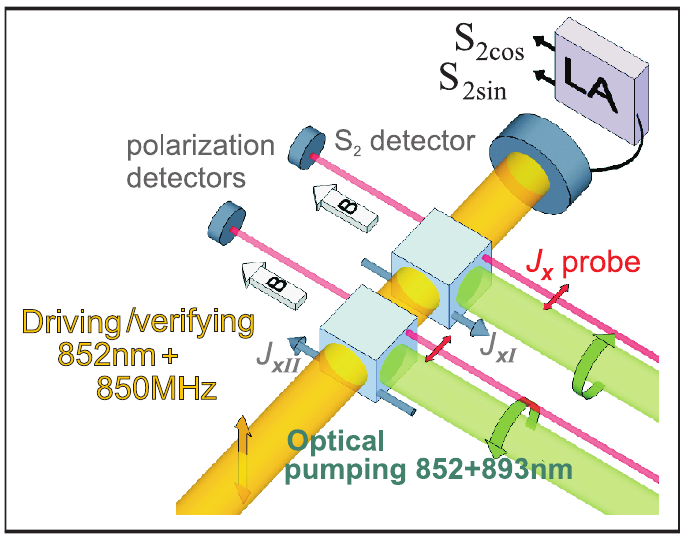}
\caption{\textbf{Measurement setup and geometry of the
experiment:} The dc polarization detectors measure the Faraday
rotation angle $\Theta$ proportional to the macroscopic spin
$J_x$. The ac $S_2$ detector signal processed by the lock-in
amplifier (LA) is used to determine the atomic quantum spin
components $J_{y,z}$ as described in the main text. }
\label{fig:calibA}
\end{figure}

Decay of the atomic state with the rate $\gamma_\text{extra}$ can
be included in the input-output equations (Eq.~(1)) as follows:
\begin{eqnarray}
(P_I^\text{out}+P_{II}^\text{out})/\sqrt{2}&=&(P_I^\text{in}+P_{II}^\text{in})/\sqrt{2}
e^{-\gamma T} -(\mu - \nu)^2\kappa{y}_{c+}^{\text{in}}\nonumber\\
&&\ \!+ \epsilon\sqrt{1-e^{-2\gamma T}}\cdot{F}_{p+}\nonumber\\
(X_I^\text{out}-X_{II}^\text{out})/\sqrt{2}&=&(X_I^\text{in}-X_{II}^\text{in})/\sqrt{2}
e^{-\gamma T} -(\mu - \nu)^2\kappa{y}_{s+}^{\text{in}}\nonumber\\
&&\ \!+ \epsilon\sqrt{1-e^{-2\gamma T}}\cdot{F}_{x-} \label{noise}
\end{eqnarray}
with the total decay
$\gamma=1/T_2=\gamma_s+\gamma_{\text{extra}}$,
$\epsilon^2=\gamma_{\text{extra}}/\gamma$ \cite{FootnoteHanna} and
the coupling constant $\kappa=\sqrt{(1-\epsilon^2)(1-e^{-2\gamma
T})}/(\mu-\nu)$. The two-cell noise operators
$\langle{F}_{i}^2\rangle=\tfrac{1}{2}$ model a decay towards the
coherent spin state (CSS), which is a good approximation on short
interaction timescales. The equations for light are adjusted
accordingly and used for the reconstruction of the atomic noise.
The coupling constant $\kappa^2$ is calibrated as discussed in
[14].  In Fig.~\ref{fig:calibB}a, measurements of the coupling
constant are shown for different atom numbers, monitored by the
Faraday angle $\Theta \propto J_x$ which is measured by the
polarization rotation of an independent probe beam propagating in
$x$-direction (see Fig.~~\ref{fig:calibA}). $(\mu-\nu)^2$ depends
on detuning and has been found experimentally [19] and derived
from the theory - both methods give the value $0.16 \pm 0.005$.
%

The imperfect detection efficiency $\eta=0.84(3)$ is included in
the input-output equations via a simple beam splitter model. The
light losses in the detection path which are relevant for the
atomic noise reconstruction  have been determined in two
independent ways. They were measured directly and also inferred
from the atomic noise as a function of the atom number presented
in Fig.~\ref{fig:calibB}b.
In order to establish the projection noise (PN) level we run a
series of experiments in which we determine $\xi(t=0)$ (more
precisely at $t=0.02$msec), i.e., right after the optical pumping
is complete as a function of the macroscopic spin $J_x$. The
number of atoms (macroscopic spins $J_x$) is varied by changing
the temperature of the cells. We use a 1ms drive pulse and perform
measurements of $S_2$ as sketched in Fig.~\ref{fig:calibA}. We
then reconstruct the atomic noise with the equations discussed
above.

We also independently measure the degree of spin polarization, as
described below, exceeding $0.992$ for all relevant atomic
numbers. The results of the atomic noise measurements
(Fig.~~\ref{fig:calibB}b) are well fit by $\xi \propto
\Theta+0.004(3) \Theta^2$. This proves that the quadratic
deviation from the PN is small and well characterized. For the
presented range of number of atoms corresponding to $\Theta =6-11$
degrees, $\xi(0)$ is a few per cent above the PN-level. We also
verify that $\xi$ is independent of the duration of the verifying
measurement $t_\text{probe}$ for 0.5ms$<t_\text{probe}<$3ms. We
conclude that
the measurement of $\xi$ is reliable within the uncertainty of
$\pm 4 \%$ arising from uncertainty in the measurements of
$\kappa^2$, the detection efficiency $\eta$, and the shot noise of
light.

%

\subsection*{Measurement of the macroscopic spin orientation}
\begin{figure}
\includegraphics[scale=0.4]{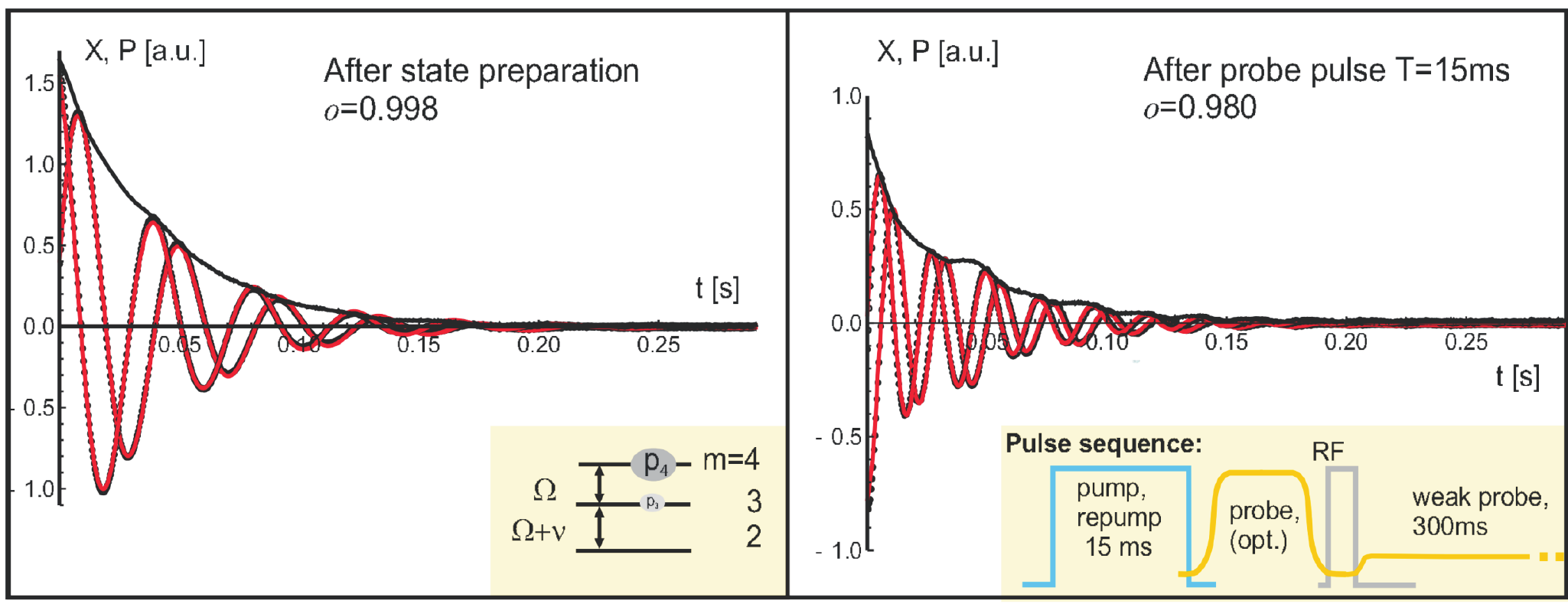}
\caption{\textbf{Orientation test} for the initial atomic state
and after 15ms probing with $5.6$mW. The time evolution of a
displacement in the two spin components in the rotating frame
$X\propto\langle J_y\rangle$ and $P\propto\langle J_z\rangle$ is
detected. The black dots are the measurement outcomes and the red
lines the fitted curves, from which the orientation $o$ can be
determined. The insets show the level structure (with $\nu\approx
20$Hz) and the pulse sequence. The first probe pulse is optional.}
\label{fig:or}
\end{figure}

The optimization of the orientation $o=\tfrac{1}{4}\sum_m m\cdot
p_m$, where $p_m$ is the population of the magnetic sub-level $m$,
is crucial for the success of the entanglement generation.
Significant imperfections of the orientation lead to atomic noise
above the PN level. Also several assumptions in the theoretical
model are only valid for a highly oriented state. In the
experiment the orientation was monitored as a function of time
during the entire probe duration using a method closely related to
the one described in \cite{Julsgaard2004}. After the preparation
of the CSS by optical pumping, we use a weak RF magnetic pulse at
the Larmor frequency to create excitations causing a displacement
of the rotating collective spin. This displacement is read out
with a weak probe beam. The coherences between the different
magnetic sub-levels oscillate at slightly different frequencies
due to the second order Zeeman shift. In our case this splitting
is 20Hz. When we look at the evolution of the displacement
demodulated with an RF frequency, close to the Larmor frequency
$\Omega$, a population of sub-levels other than $m=4$ will
manifest itself in a quantum beat signal at multiples of the
second order Zeeman splitting frequency. In Fig.~\ref{fig:or},
two signals for slightly different orientations are shown and the
difference is clearly visible. This method is remarkably sensitive
to orientation imperfections. After the pumping, we start very
close to the CSS with $o=0.998(3)$. After a 15ms probe pulse with
5.6mW the orientation is reduced to $o=0.980(3)$.
%
%
%
\section*{Comparison of experimental data and theoretical predictions}
The generated amount of entanglement $\xi(t)$ can be calculated as
described in detail in [12]. As explained there, the produced
entanglement is given by
\begin{eqnarray}\label{ExpEnt}
\xi(t)=\frac{\Sigma_{J}(t)+14N_{|4,\pm3
\rangle}(t)}{N_2(t)\left(P_{2}(t)+7\right)},
\end{eqnarray}
where $N_{|4,\pm3 \rangle}$ ($N_{\uparrow/\downarrow}$) is the
number of particles in state $|4,\pm3 \rangle$
($|\!\uparrow\rangle/|\!\downarrow\rangle$) and
$N_2=N_{\uparrow}+N_{\downarrow}$ and
$P_2=\left|N_{\uparrow}-N_{\downarrow}\right|/N_2$ denote the
particle number and orientation with respect to the two-level
subsystem respectively. The time evolution of the EPR variance
$\Sigma_{J}=\text{var}\left(J_{y,I}-J_{y,II}\right)+\text{var}\left(J_{z,I}-J_{z,II}\right)$
(see main text) can be calculated using Eq.~(21) in [12] if the
collective decay rate ($d\Gamma$) and the effective dephasing rate
($\tilde{\Gamma}$) as well as $N_2(t)$ and $P_2(t)$ are known. In
the following we describe how these quantities can be inferred and
explain the theoretical fits to the measured data
presented in the main text.\\
\\The essential features of the experiment can be described by
means of a simplified model, which allows one to take additional
dynamics due to the multilevel character of Cesium into account
and involves only the three atomic states $|4,\pm 4\rangle$,
$|4,\pm 3\rangle$ and $|h\rangle_{I/II}\equiv|3,\pm 3\rangle$. For
the timescales considered here, atomic population in other
sublevels can be neglected. Since the model is primarily intended
to describe qualitatively the physical effects observed in the
experiment with very few parameters, we use
$\Gamma_{|4,\pm4\rangle \rightarrow|h\rangle}\approx
\Gamma_{|4,\pm3\rangle \rightarrow|h\rangle}=\Gamma_{\rm{out}}$
and $\Gamma_{|h\rangle\rightarrow |4,\pm4\rangle}\approx
\Gamma_{|h\rangle \rightarrow|4,\pm3\rangle}=\Gamma_{\rm{in}}$
such that
\begin{eqnarray*}
\frac{d}{dt}\ \! N_2(t)&=&-\left(\Gamma_{\rm{out}}+2\Gamma_{\rm{in}}\right)N_2(t)+2N\Gamma_{\rm{in}},\\
\frac{d}{dt}\ \!
\tilde{P}_2(t)&=&-\left(\Gamma_{3,4}+\Gamma_{4,3}+\Gamma_{\rm{out}}\right)\tilde{P}_2(t)\\
&&+\left(\Gamma_{3,4}-\Gamma_{4,3}\right)N_2(t)/N,
\end{eqnarray*}
where $\tilde{P}_2(t)\!=\!P_2(t)N_2(t)/N$. Here and in the
following we use the abbreviations $\Gamma_{|4,\pm
4\rangle\rightarrow|4,\pm 3\rangle}\!=\!\Gamma_{4,3}$ and
$\Gamma_{|4,\pm 3\rangle\rightarrow|3,\pm
4\rangle}\!=\!\Gamma_{3,4}$.
Atomic transitions can be either induced by the driving field or
due to collisions. Since the thermal energy of atoms is much
larger than the atomic level splittings, we assume the same
collisional rate $\Gamma_{\text{col}}$ for all atomic transitions.
Accordingly,
\begin{eqnarray*}
\Gamma_{3,4}&=&\mu^2\ \Gamma+\Gamma_{\text{col}},\ \ \ \ \
\Gamma_{\text{out}}=\Gamma_{\text{L}}^{\text{out}}+\Gamma_{\text{col}},\\
\Gamma_{4,3}&=&\nu^2\ \Gamma+\Gamma_{\text{col}},\ \ \ \ \ \ \ \!
\Gamma_{\text{in}}=\Gamma_{\text{col}},
\end{eqnarray*}
where the abbreviations $\Gamma_{|4,\pm 4\rangle\rightarrow|4,\pm
3\rangle}=\Gamma_{4,3}$ and $\Gamma_{|4,\pm
3\rangle\rightarrow|3,\pm 4\rangle}=\Gamma_{3,4}$ have been used.
$\mu^2\Gamma$ and $\nu^2\Gamma$ are the driving field induced
cooling and heating rate respectively.
$\Gamma_{\text{L}}^{\text{out}}$ is the rate at which atoms leave
the two level subsystem due to radiative transitions caused by the
driving field.
The number of free parameters in these equations can be reduced to
two, using the experimentally determined time derivative of the
atomic polarization $P=\langle J_x(t)\rangle/\langle
J_x(0)\rangle$ at time $t=0$
\begin{eqnarray*}
\frac{d}{dt}\!P(t)\bigg|_{\tiny{t=0}}\!\!\!\!\!\!\!&=&\!\!\!\!\frac{-\left(\Gamma_{4,3}\!+\!4\Gamma_{\rm{out}}\right)\!N_{|4,\pm
4\rangle}(0)\!+\!\left(\Gamma_{3,4}\!-\!3\Gamma_{\text{out}}\right)\!N_{|4,\pm
3\rangle}(0)}{\langle J_x(0)\rangle},
\end{eqnarray*}
where it is taken into account that the initial spin state is not
perfectly polarized, but contains a small fraction of atoms in
state $|4,\pm 3\rangle$.
The initial populations $N_{|4,\pm 4\rangle}(0)=0.99$, $N_{|4,\pm
3\rangle}(0)=0.01$ and $N_{\text{h}}(0)=0$ are estimated based on
measurements of the orientation of the initial spin state after
optical pumping which are described in the first part of SM.
Using this constraint, $P(t)$ can be fitted with two free
parameters. This way, fixed expressions for $P_2(t)$ and $N_2(t)$
are obtained. As mentioned above, the values of the collective
decay rate $d\Gamma$ and the dephasing rate due to noise effects
$\tilde{\Gamma}$ \cite{FootnoteChristine} have to be known in
order to calculate the generated amount of entanglement as
described in [12]. These parameters are determined from the
experimentally obtained slope of the variance
$\Sigma_{J}(t)/\left(2|\langle J_x\rangle|\right)$ at time $t=0$
\begin{eqnarray*}
\frac{d}{dt} \frac{\Sigma_J(t)}{2|\langle J_x(0)\rangle|}
\bigg|_{t=0}\!\!\!\!\!&=& \!\!\!-4 Nd\Gamma
P_2(0)\!\left(1\!-\!P_2(0)/\!\left(\mu\!-\!\nu\right)^2\right)\\
\!\!\!\!\!&&+7\Gamma_{\text{in}}P_2(0)\!-\!7\left(\Gamma_{\text{out}}\!+\!\Gamma_{3,4}\!-\!\Gamma_{4,3}\right)\!N_{|4,\pm
3\rangle}(0),
\end{eqnarray*}
and the decay of the transverse spin
$ \langle
J_y(t)\rangle=e^{-\frac{1}{2}\left(\tilde{\Gamma}+d\Gamma
\tilde{P}_2(t)\right)t}\langle J_y(0)\rangle$,
where Eq.~(21) in [12] and the identities relating quantities
defined with respect to a two-level system to quantities defined
with respect to a multi-level structure presented in Sec.~IV.A in
[12] have been used.

\begin{thebibliography}{54}
\bibitem{Footnote} The effective dephasing rate associated with noise effects
$\tilde{\Gamma}=\Gamma_{\text{cool}}+\Gamma_{\text{heat}}+\Gamma_{\text{deph}}$
includes the effective single particle cooling rate
$\Gamma_{\text{cool}}$ (which is a sum of rates corresponding to
cooling processes due to the driving and pump field and
collisions), the effective single particle heating rate
$\Gamma_{\text{heat}}=\Gamma_{\uparrow\downarrow}=\nu^2\Gamma+\gamma_{\text{\text{col}}}$
as well as pure dephasing $\Gamma_{\text{deph}}$.
\end{thebibliography}
\begin{thebibliography}{1}
\bibitem[S1]{FootnoteHanna}
In the actual experiment, the $\gamma_{\text{extra}}$ will collect
all decay mechanisms coming from spontaneous emission, collisions
and magnetic field inhomogeneity
\bibitem[S3]{FootnoteChristine}
In the absence of pump or repump fields, the dephasing rate
associated with noise effects
$\tilde{\Gamma}=\Gamma_{3,4}+\Gamma_{4,3}+\Gamma_{\text{deph}}$
includes the effective single particle cooling rate
$\Gamma_{3,4}=\mu^2 \Gamma+\Gamma_{\text{col}}$, which is a sum of
the rates corresponding to cooling processes due to the driving
field and collisions, the effective single particle heating rate
$\Gamma_{4,3}=\nu^2 \Gamma+\Gamma_{\text{col}}$ as well as pure
dephasing $\Gamma_{\text{deph}}$.
\end{thebibliography}
\end{document}